\documentclass[12pt]{oxarticle}
\pdfoutput=1

\usepackage[hypertexnames=false,linktocpage=true]{hyperref}
\hypersetup{colorlinks=true, citecolor=blue, linkcolor=blue, urlcolor=blue, pdfauthor={}, pdftitle={},breaklinks=true}

\usepackage{graphicx, float, array, xspace, amscd, amsmath, amsthm, amssymb, latexsym, relsize, bbm}
\usepackage[left=1.7in, right=0.3in, top=0.8in, bottom=1.2in]{geometry}
\usepackage[bbgreekl]{mathbbol}
\usepackage{amsfonts, tikz-cd}
\usepackage[T1]{fontenc}
\usepackage{tikz}
\usetikzlibrary{backgrounds}
\usepackage[framemethod=tikz]{mdframed}
\AtBeginEnvironment{mdframed}{%
\tikzset{every picture/.style={}}%
}
\mdfsetup{roundcorner=.5ex}
{\end{mdframed}}

\DeclareSymbolFontAlphabet{\mathbb}{AMSb} 
\DeclareSymbolFontAlphabet{\mathbbl}{bbold}

\usepackage[sort&compress, comma, square, numbers]{natbib}
\usepackage[all,cmtip]{xy}
\usepackage{color}
\definecolor{MyDarkBlue}{rgb}{0.15,0.25,0.45}

\let\SS=\S 

\newcommand{\Dch}{{\check{D}}}

\newcommand{\chD}{{\check{D}}}

\renewcommand{\#}{^{\sharp}}

\newcommand{\LC}{\text{\tiny LC}}

\renewcommand{\sb}{{\overline{\sigma}}}

\newcommand{\rb}{{\overline{ r}}}

\newcommand{\w}{{\,\wedge\,}}
\newcommand{\wt}{\widetilde}

\newcommand{\pre}[1]{{}^{(#1)}}

\newcommand{\Le}{{\mathfrak{e}}}

\newcommand{\Lg}{{\mathfrak{g}}}

\newcommand{\fD}{{\mathfrak{D}}}

\newcommand{\delslash}{\ensuremath \raisebox{0.025cm}{\slash}\hspace{-0.23cm} \del}
\newcommand{\Hslash}{\hspace{0.1cm}\ensuremath \raisebox{0.03cm}{\slash}\hspace{-0.30cm} H}

\newcommand{\Fslash}{\hspace{0.1cm}\ensuremath \raisebox{0.025cm}{\slash}\hspace{-0.28cm} F}

\newcommand{\half}{\frac{1}{2}}
\newcommand{\qrt}{\frac{1}{4}}

\def\rep#1{{{\boldsymbol{#1}}}}
\def\brep#1{{{\overline{\boldsymbol{#1}}}}}

\newcommand{\ab}{{\overline\alpha}}
\newcommand{\bb}{{\overline\beta}}

\renewcommand{\aa}{\mathfrak{a}}

\renewcommand{\a}{\alpha}

\newcommand{\G}{\Gamma}
\renewcommand{\d}{\delta}\newcommand{\D}{\Delta}
\newcommand{\ve}{\varepsilon}

\newcommand{\Th}{\Theta}

\renewcommand{\k}{\kappa}
\renewcommand{\L}{\Lambda}
\newcommand{\m}{\mu}
\newcommand{\n}{\nu}

\renewcommand{\r}{\rho}
\newcommand{\s}{\sigma}\renewcommand{\S}{\Sigma}

\newcommand{\vph}{\varphi}

\DeclareFontFamily{OT1}{pzc}{}
\DeclareFontShape{OT1}{pzc}{m}{it}{<-> s * [1.200] pzcmi7t}{}
\DeclareMathAlphabet{\mathpzc}{OT1}{pzc}{m}{it}

\newcommand{\ccB}{\mathpzc B}

\newcommand{\cD}{\mathcal{D}}

\newcommand{\cF}{\mathcal{F}}

\newcommand{\cL}{\mathcal{L}}
\newcommand{\cM}{\mathcal{M}}

\newcommand{\cO}{\mathcal{O}}

\newcommand{\cR}{\mathcal{R}}

\newcommand{\ccY}{\mathpzc Y}
\newcommand{\ccZ}{\mathpzc Z}

\newcommand{\ccZb}{{\overline \ccZ}}

\DeclareFontFamily{U}{bbold}{}
\DeclareFontShape{U}{bbold}{m}{n}
 {  <-5.5> s*[1.05] bbold5
    <5.5-6.5> s*[1.05] bbold6
    <6.5-7.5> s*[1.05] bbold7
    <7.5-8.5> s*[1.05] bbold8
    <8.5-9.5> s*[1.05] bbold9
    <9.5-11.5> s*[1.05] bbold10
    <11.5-16> s*[1.05] bbold12
    <16-> s*[1.05] bbold17
 }{}

\newcommand{\IR}{\mathbbl{R}}

\font\eightrm=cmr8 at 8pt
\font\csc=cmcsc10

\newcommand{\beq}{\begin{equation}}
\newcommand{\eeq}{\end{equation}}
\newcommand{\beqnn}{\begin{equation*}}
\newcommand{\eeqnn}{\end{equation*}}
\newcommand{\bea}{\begin{eqnarray}}
\newcommand{\eea}{\end{eqnarray}}
\newcommand{\bean}{\begin{eqnarray*}}
\newcommand{\eean}{\end{eqnarray*}}

\newcommand{\sref}[1]{\SS\ref{#1}}

\newcommand{\place}[3]{\vbox to0pt{\kern-\parskip\kern-7pt
                             \kern-#2truein\hbox{\kern#1truein #3}
                             \vss}\nointerlineskip}

\DeclareFontFamily{U}{wncy}{}
\DeclareFontShape{U}{wncy}{m}{n}{<->wncyr10}{}
\DeclareSymbolFont{mcy}{U}{wncy}{m}{n}
\DeclareMathSymbol{\sha}{\mathord}{mcy}{"58}

\newcommand{\eu}[1]{{\mathfrak #1}}

\newcommand{\del}{{\partial}}
\newcommand{\delb}{{\overline{\partial}}}

\newcommand{\nb}{{\overline\n}}
\newcommand{\mb}{{\overline\m}}

\newcommand{\Dbar}{{\overline D}}

\renewcommand{\aa}{\mathfrak{a}}

\newcommand{\dd}{{\text{d}}}

\newcommand{\ddH}{{\dd_{\nabla^\H}}}

\renewcommand{\H}{\text{H}}

\newcommand{\vol}{\mbox{\,vol}}
\newcommand{\tr}{\text{Tr}\hskip2pt}

\newcommand{\ap}{{\a^{\backprime}\,}}
\renewcommand{\sb}{{\overline{\sigma}}}

\renewcommand{\rb}{{\overline{\rho}}}
\renewcommand{\=}{\;=\;}

\makeatletter
\g@addto@macro\bfseries{\boldmath}
\makeatother

\newcommand{\citeE}{\cite{McOrist:2021dnd}\xspace}
\newcommand{\citeES}{\cite{McOrist:2024zdz}\xspace}
\newcommand{\citeM}{\cite{Candelas:2016usb}\xspace}

\newcommand{\citeSG}{\cite{McOrist:2019mxh}\xspace}

\newcommand{\citeUG}{\cite{Candelas:2018lib, McOrist:2019mxh}\xspace}

\newcommand{\citeUS}{\cite{Candelas:2018lib, Candelas:2016usb,McOrist:2016cfl}\xspace}

\newcommand{\citeAQS}{\cite{Anguelova:2010ed}\xspace}

\newcommand{\citeMSS}{\cite{McOrist:2024ivz}\xspace}
\newcommand{\citeMSSUG}{\cite{McOrist:2024glz}\xspace}

\renewcommand{\baselinestretch}{1.1}
\numberwithin{equation}{section}
\proofmodefalse
\begin{document}
\pagestyle{empty}      
\ifproofmode\underline{\underline{\Large Working notes. Not for circulation.}}\else{}\fi
\hypersetup{pageanchor=false}

\begin{center}
\null\vskip-0.5in
{\Huge  The heterotic $G_2$ moduli space metric \\[0.3in]}
{\csc   Jock McOrist$^{\sharp 1}$,  Martin Sticka$^{\sharp 2}$ and Eirik Eik Svanes$^{\dagger 3}$\\[0.3in]}

{\it 
$^\sharp$ Department of Mathematics\hphantom{$^2$}\\
School of Science and Technology\\
University of New England\\
Armidale, 2351, Australia\\[3ex]
}

{\it 
$^\dagger$ Department of Mathematics and Physics\hphantom{$^1$}\\
Faculty of Science and Technology\\
University of Stavanger\\
N-4036, Stavanger, Norway\\[3ex]
}

\footnotetext[1]{{\tt jmcorist@une.edu.au}}
\footnotetext[2]{{\tt msticka@myune.edu.au}}
\footnotetext[3]{{\tt eirik.e.svanes@uis.no}}
\vspace{1cm}
{\bf Abstract\\[-15pt]}
\end{center}

In this article we dimensionally reduce a heterotic supergravity on a $G_2$ background with Minkowski spacetime using a certain cohomology as a basis for the Kaluza-Klein expansion, up to and including first order in $\ap$. We construct the moduli space heterotic $G_2$ compactifications. The $\ap$-correction induces a curvature correction to the Weyl-Peterson metric. In the limit in which the $G_2$ manifold reduces to $SU(3)$, we recover known results.

\vskip150pt

\newgeometry{left=1.5in, right=0.5in, top=0.75in, bottom=0.8in}
%
\newpage
%
%
%
\setcounter{page}{1}
\pagestyle{plain}
\hypersetup{pageanchor=true}
\renewcommand{\baselinestretch}{1.3}
\null\vskip-10pt

\section{Introduction}

 A heterotic compactification on a $G_2$-manifold $Y$ realises a three-dimensional spacetime $\IR^{2,1}$ or $AdS_3$. Such vacua  were first studied in the context of heterotic compactifications in  \cite{Gunaydin:1995ku,Gauntlett:2001ur, Friedrich:2001nh, Friedrich:2001yp, Gauntlett:2002sc, Gauntlett:2003cy, Ivanov:2003nd, Ivanov:2009rh, Kunitomo:2009mx, Lukas:2010mf, Gray:2012md}. We aim to explore the metric structure of the moduli space of heterotic compactifications preserving $\IR^{2,1}$ spacetimes. This is the manifold representing parameters for unobstructed deformations of the equations of motion. All deformations considered preserve supersymmetry, as we currently lack the theoretical tools to handle supersymmetry-breaking cases, though they are of great physical interest. Our main approach uses supergravity as an approximation to string theory. This is a formal expansion in a small parameter $\ap \ll 1$, with increasingly complex differential equations emerging order-by-order in $\ap$. The fields are similarly expanded, and the system is solved iteratively. We work to first order in $\ap$, where much is already known. At this order, the geometric data defining a solution includes the $G_2$ manifold $Y$, the $G_2$ structure $\varphi$, a compatible metric $g$ on $Y$, a bundle $V \to Y$ with a connection $A$ embedded in $E_8 \times E_8$, and a three-form $H$.
 The supersymmetry conditions imply $F$ satisfies an instanton condition \cite{Gauntlett:2002sc, Gauntlett:2003cy,Ivanov:2003nd,ReyesCarrion:1998si}:
$
F \w \psi \= 0~.
$
The modifications to this set of equations and their solutions at higher orders in the $\ap$ expansion are of great interest. Although we don’t know the exact form of these corrections, it is essential that they remain smaller than the first-order results discussed here. 

We require  the $G_2$ structure $\varphi$  be integrable \cite{fernandez1998dolbeault, Gauntlett:2001ur, Friedrich:2001yp}  so that there is   a unique metric connection on the tangent bundle $TY$ with totally antisymmetric torsion \cite{Friedrich:2001nh}, we label $\Th^B$. There are two other  relevant connections: the Hull connection $\Th^\H$ which has completely antisymmetric torsion of opposite sign to $\Th^B$, and the Levi-Civita connection $\Th^\LC$. We denote these curvatures $R^\H$, $R^B$, and  $R$ for Levi-Civita. These connections are all determined in terms of the underlying metric and three-form $H$.  
This data satisfies a Bianchi identity that derives from the Green-Schwarz anomaly cancellation condition on the worldsheet:
\beq
\dd H \=\!- \frac{\ap}{4} \Big( \tr (F\w F) - \tr ( R^\H\w R^\H)\, \Big)~.
\label{eq:Anomaly0}
\eeq
The connection used in the Chern class $\tr R^\H \wedge R^\H$ is chosen to match string theory scattering amplitudes and is also favoured for its compatibility between the supergravity equations of motion and supersymmetry. A result of this paper is to observe that the Hull connection is crucial for the moduli space metrics between $G_2$ and $SU(3)$ compactifications to be compatible. Hence, it is the moduli space defined by this connection within string theory that we seek to study.

The moduli space of this datum has been studied recently in \cite{delaOssa:2016ivz, Clarke:2016qtg, delaOssa:2017pqy, delaOssa:2017gjq, Fiset:2017auc, Clarke:2020erl, Silva:2024fvl, delaOssa:2024dzo}. In \cite{Clarke:2016qtg} it was proven in the mathematics literature this space of deformation is finite dimensional (up to first order in $\ap$), which is important given the lack of physical evidence analogous to say \cite{Nemeschansky:1986yx,Jardine:2018sft} and \cite{Witten:1985bz,Witten:1986kg}  who give worldsheet and spacetime finiteness arguments respetively for $d=4$ $SU(3)$ compactifications. The role of $\ap$--corrections is interesting even for $G_2$ holonomy manifolds as pointed out in \cite{Becker:2014rea}, in which it is not clear if the vacua exist due to the low amount of spacetime supersymmetry ($N=1$, $d=3$) and due to the lack of an argument following \cite{Nemeschansky:1986yx}. It could be the case, for example, that perturbative or non-perturbative $\ap$-corrections ruin the vacuum cause fields to run away to flat spacetime. A possible route for constraining the $\ap$-corrections from a spacetime perspective is to study the analogue of the universal bundle \citeUG for $G_2$ compactifications and its moduli \citeMSSUG. Setting these important but difficult issues aside, we work under the assumption the vacuum exists and is embeddable in string theory. Given that context, recently explicit solutions with $AdS_3$ spacetimes have been constructed in \cite{delaOssa:2021qlt, Lotay:2021eog, Galdeano:2024fsc}. One approach to studying the space of deformations of these theories is inspired by \cite{Atiyah:1955} and construct a complex of vector bundles via a nilpotent operator $D$ whose nil-potency relies explicitly on the representation theory of $G_2$, as well as the  conditions for supersymmetry to first order in $\ap$ and the Bianchi identities for the curvatures $R^\H, F$ and $H$. We can do this because integrable $G_2$ structures have features in common with complex geometry and facilitate  a canonical differential complex $\Lambda^*(Y)$ with associated cohomologies $H^*(Y)$. These have similarities with the Dolbeault complex of complex geometry. Families of $SU(3)$ structure manifolds can be examined through an embedding in integrable $G_2$ geometry, see e.g. \cite{delaOssa:2014lma}. This embedding allows variations of complex and Hermitian structures of six-dimensional manifolds to be considered equivalently. It is important to note that we do not expect a self-consistent truncation of heterotic string theory to first order in $\ap$. That being so, we are working perturbatively in an aysmpototic expansion in $\ap$ (and of course $g_s$), and all the data described above is well-defined only up to $\ap^2$ corrections. This includes the complex defined by $D$: the supersymmetry variations have  (to date uncomputed) $\ap^2, \ap^3, \cdots$ corrections and so the definition of $D$ is also suppressing these corrections. In physics this is perfectly well-defined thing to do as we are working semi-classically, and are in a similar situation to computing perturbative expansions of operators in quantum field theory. 

In \cite{delaOssa:2017pqy, Clarke:2020erl} the moduli space of the supersymmetric $G_2$ solutions was studied by modifying the string theory Bianchi identity \eqref{eq:Anomaly0} so that the connection used to compute $\tr R^2$ is an instanton with respect to the $G_2$ structure. While the Hull connection is an instanton to zeroth order in $\ap$, the instanton connection carries with it additional non-physical deformations. These are spurious directions in the moduli space and are to be solved for in terms of the physical deformations.\footnote{The result of \cite{Clarke:2020erl} shows this moduli problem is elliptic, resulting in a finite-dimensional moduli space. The physical moduli space is a sub-space of this moduli space.} This was done in the context of $SU(3)$ \cite{McOrist:2021dnd, McOrist:2024zdz, deLazari:2024zkg}, and inpsired by that approach a recent paper \citeMSS achieved this for $G_2$ compactifications. This gave rise to an operator $\Dch$ and associated complex defining the physical moduli, in addition to a conjecture for the $G_2$ moduli space metric.  Two non-trivial checks were provided: first the metric reduced to $SU(3)$ to the metric constructed in \citeUS; secondly the metric defined an adjoint $\Dch^\dag$ and the kernel of this adjoint both provided a natural gauge fixing for the $G_2$ moduli and furthermore reduced to precisely the D-terms constructed in \citeE. This led to the pretty result that the $G_2$ moduli are described by the intersection of two kernels:\footnote{As we want this to be describing heterotic string theory, we are working in the effective field theory approximation, which is an asymptotic expansion in two small parameters $\ap,g_s$, both of which are taken to be small. We work to zeroth order in $g_s$ and to first order in $\ap$. All equalities are understood to be modulo $\ap^2$-corrections. At higher orders in both these quantities both  operators $\Dch$, its adjoint $\Dch^\dag$, the field $\ccY$ may receive corrections, but these are yet to be computed.}
\beq
\Dch \ccY \= 0~, \qquad \Dch^\dag \ccY \= 0~.
\eeq
This is the to be compared to the result in \citeE: 
\beq
\Dbar \ccY \=0~, \qquad \Dbar^\dag \ccY \= 0~.
\eeq
they look formally very similar but their interpretations are different. The $SU(3)$ equations correspond to F-terms and D-terms respectively; the $G_2$ equations correspond to the moduli being in a cohomology and the fixing of gauge symmetries. Nonetheless, both show that if the moduli are viewed in the context of the right sort of vector bundle and cohomology they correspond to harmonic representatives. While this is a natural result it did not have to be this way! For example, when studying $SU(3)$ moduli and the fixing of gauge symmetries in \citeSG, the moduli fields or space of deformations are certainly not harmonic when viewed with respect to $\delb$-cohomology.
 
The first goal of this paper is modelled on the calculation in \citeM: we are to construct the $G_2$ moduli space metric based on a dimensional reduction. Unlike \citeM, in this case we are in the fortunate position of having a conjectured form to check against. We will nonetheless do the calculation and show that the metric in \citeMSS is the $G_2$ metric. It is qualitatively important that (a) the metric received $\ap$-corrections already at first order: one involving the vector bundle and one scaling like the Riemann curvature of the $G_2$ metric, and (b) the spurious modes are eliminated by setting the spin connection to be the Hull connection.

\section{ \texorpdfstring{$G_2$}{G2} compactifications of heterotic string theory}
We adopt the same notation for connections, torsion and tensors used in \citeMSS and \citeM. 

\subsection{Ten-dimensional supergravity with first order \texorpdfstring{$\ap$}{ap}: conventions}
In the field basis of \cite{Bergshoeff:1989de,Bergshoeff:1988nn} the ten--dimensional $\ap^2$--supergravity theory can be written down explicitly. The action is
\begin{equation}
S \= \frac{1}{2\kappa_{10}^2} \int\! \dd^{10\,}\! X \sqrt{g_{10}}\, e^{-2\Phi} \Big\{ \cR -
\half |H|^2  + 4(\del \Phi)^2 - \frac{\ap}{4}\big( \tr |F|^2 {-} \tr |R^\H|^2 \big) \Big\} + \cO(\ap^3)~.
\label{eq:10daction}
\end{equation}
On the $G_2$ manifold $Y$, we denote $m,n,\ldots$ real indices along $Y$.\footnote{The 10D Newton constant is denoted by $\kappa_{10}$,
\hbox{$g_{10}=-\det(g_{MN})$}, $\Phi$ is the 10D dilaton.} $\cR$ is the Ricci scalar evaluated using the Levi-Civita connection and $F$ is the Yang--Mills field strength with the trace taken in the adjoint of the gauge group. We take the generators of the gauge group $T^a$ and the trace $\tr$ to be defined so that  $\tr (T^a T^b)$ is positive definite. We will often need wedge products involving the                        basis elements of $T^*Y$ written in a coordinate basis. To save writing excessive wedges we will write $\dd x^{m_1\cdots m_n} \cong \dd x^{m_1} \w \cdots \wedge \dd x^{m_n}$. If any possible confusion arises we will use the $\wedge$ symbol explicitly. 

There is a point-wise inner product  on $p$-forms given by
$$
\langle S,\, T\rangle~=~ 
\frac{1}{p!} \, g^{M_1 N_1} \ldots g^{M_p N_p}\, S_{M_1\ldots M_p} \,T_{N_1 \ldots N_p}~, \qquad |T|^2 ~=~\langle T,\, T\rangle~.
$$
 The three--form $H$ satisfies a Bianchi--identity
\beq
\dd H =- \frac{\ap}{4} \left( \tr F\w F - \tr R^\H\w R^\H \, \right)~.
\notag\eeq

The 10-dimensional background being supersymmetric amounts to the vanishing gravitino, dilatino and gaugino variations:
\beq
\begin{split}
 \d \Psi_M &\= \nabla_M^B \ve  + \cO(\ap^2) \= 0 ~,\\
 \d \lambda &\=\! - \frac{1}{2\sqrt{2}} \left( \delslash - \half \Hslash  \right) \ve  + \cO(\ap^2) \= 0~,\\
 \d \chi &\=\! - \frac{1}{2} \Fslash \ve + \cO(\ap^2) \= 0~,
\end{split}
\label{eq:10dsusy}
\eeq
where we use the usual slash notation. We have written the formal expansion parameter $\ap^2$ explicitly; there are equations not written to do with second order equations, beyond our remit. The important point is that the spinor $\ve$ is covariantly constant with  torsion opposite to what appears in the action, Bianchi identity and equations of motion. 

Let us assume $H = \cO(\ap)$ so that  $Y$ is $G_2$ holonomy to zeroth order in $\ap$ and the corresponding curvature is Ricci-flat. This is  a physical  requirement -- if we take $H=\cO(\ap)$, it is guaranteed that the $G_2$ geometry will define a weakly coupled worldsheet sigma model that underpins a string theory vacuum.

\subsection{\texorpdfstring{$G_2$}{G2} complex}
As a first step we recall the complex constructed in \citeMSS. The bundle $Q$ is topologically 
$$
Q\= T^*Y \oplus {\rm End}(V)~.
$$
A basis of sections  are denoted by
$$
\ccY_a \= 
\begin{pmatrix}
 M_a \\
 \aa_a
\end{pmatrix}~, 
$$
where $a=1,\cdots, n$ with $n=\dim \cM_{G_2}$ runs over the dimensions of the moduli space. 

We will need to study $p$-forms valued in $Q$ and by an abuse of notation use the same symbol for such a section: $\ccY\in \L^p (Y, Q)$. In calculations we will have need for indices, and the order of the indices is important as they are not democratic. For example, $M \cong M_{P m} \dd x^P \otimes \dd x^m$, where $P$ is a multi-index of degree $d$ with $\dd x^P = \dd x^{p_1\cdots p_d}/d!$. Very often $d=1$ and so $M$ is a 1-form valued in the cotangent bundle $T^*Y$.\footnote{This is not strictly correct as $M$ does not transform under symmetries as a section of the cotangent bundle, instead the $B$-field has a twist with the gauge field resulting in a more complicated transformation law. See \citeMSS. However, for the purposes of keeping track of indices, we write $T^*Y$.} We sometimes explicitly write the index corresponding to the cotangent bundle viz. $M_m\cong M_{Pm} \dd x^P$.

We will need later that $K$ is a deformation of the Hull connection in spinor indices. This is computed in \citeMSS first by going to the coordinate frame and evaluating:
\beq\label{eq:defOfHull}
\d \G^\H_q{}^s{}_t \= \d \G_q{}^s{}_t + \half \d H_q{}^s{}_t ~.
\eeq
A deformation of the Levi-Civita symbol $\G$ and three-form $H$ is
\beq\label{eq:deltaH}
\begin{split}
\d \G_q{}^s{}_t &\= g^{sp} \left( \nabla_{[t} \d g_{p]q} + \half \nabla_q \d g_{tp} \right)~,\\
\d H_q{}^s{}_t &\= g^{sm} \d g_{mn} g^{np} H_{qtp} + g^{sp}\left( \nabla_q \ccB_{pt} + \nabla_t \ccB_{qp} + \nabla_p \ccB_{tq} \right)~,
\end{split}
\eeq
where 
$$
\d H \=  \dd \ccB - \frac{\ap}{2} \tr \big( \d A F \big)~,
$$
 with $\ccB$ a gauge invariant combination of the $B$-field and gauge connection 
 $$
 \ccB  \=  \d B +\frac{\ap}{4}\tr{(A\,\d A)} + \dd \textrm{-closed}~.
 $$
Setting 
\beq\label{eq:M}
M_{ab} = \frac{1}{2} (\d g_{ab} + \ccB_{ab})~, 
\eeq
we identify
\beq\label{eq:KinM}
\d \G^\H{}^s{}_t \= \left(  \dd_\nabla M^s{}_t+ \nabla_t M{}^s - \nabla^s M_{t} \right) ~,
\eeq
where $\d \G = \d \G^\LC + \half \d H$ is a deformation of the Hull connection in coordinate indices. Note that if we lower the $s$ index, this is not antisymmetric in $s,t$ indices, as is required for metric compatibility of the spin connection; or for it to be valued in endomorphisms of $T_Y$. For this to be the case, we need to perform a transformation so the quantity is a deformation of the spin connection. This is spelled out in Appendix \sref{s:AppSpin} using results and notation from say \cite{Eguchi:1980jx}. From  \eqref{eq:ThH2} we find 
\beq\label{eq:dThH}
K_{ab} \= \d \Th^\H_{ab}  \=- E_a{}^s E_b{}^t \left(  \nabla_s M_t - \nabla_t M_{s} \right)~,
\eeq
where $e^a = e^a_q \dd x^q$ and $E_e = E_a^q \del_q$ are the vielbein and its inverse with $a,b$ flat indices. 
We have identified $K_{ab} =\d \Th^\H_{ab}$. For us $K$ always appears in quantities multiplied by an $\ap$ and so we can work to zeroth order in $\ap$. Hence, as $H=\cO(\ap)$,  $\nabla$ acts as Levi-Civita on all indices.

The bundle $Q$ is acted upon by an operator
\beq\label{eq:Dtilde}
 \cD \= 
 \begin{pmatrix}
 \ddH + R \nabla   & - \cF_1\\
 \cF_2  & \dd_A
\end{pmatrix}~,
\eeq
where 
\beq\label{eq:defnRnabcF}
\begin{split}
 (R\nabla)(M_m)& \= - \frac{\ap}{2} \, R_{qm}{}^t{}_s \nabla_t M_P{}^s \dd x^q\w \dd x^P~,  \\
\cF_1(\aa)_m &\=   \frac{\ap}{4} \tr (  F_{mq} \dd x^q \w \aa)~,\qquad \cF_2(M) \=  g^{mn}  F_{nq} M_{Pm} \dd x^q\w \dd x^P ~.
\end{split}
\eeq
It can be checked that $\chD^2 = 0$ \citeMSS. 
Acting with \eqref{eq:Dtilde} on $\ccY$  we have
\beq\label{eq:DY}
\cD \ccY \= 
\begin{pmatrix}
\dd_\nabla^\H  M_{m} - \frac{\ap}{2} R_m{}^{st} \nabla_s  M_t + \frac{\ap}{4} \tr \left( \aa F_m \right)  \\[5pt]
 -  M_{ m} F^m + \dd_A \aa
\end{pmatrix}~.
\eeq
Recall $F_m = F_{mq} \dd x^q$, $R_m{}^t{}_s = R_{qm}{}^t{}_s \dd x^q$, and $M^s = M_q{}^s \dd x^q$.

We need to gauge fix to describe the physical moduli. There is a preferred gauge fixing so that the fields reduce to the $SU(3)$ moduli calculation in \citeM, \citeE, \citeES (holomorphic gauge in \citeSG):
\beq\label{eq:gaugefix} 
\nabla^{\H\, q}M_{qm} - \frac{\ap}{2} R_m{}^{spt}  (\nabla_t M_{sp})  - \frac{\ap}{4} \tr  F_{ms} \aa^s \= 0~,
\qquad \dd_A^\dag \aa + F^{qm}  M_{qm} \= 0~.
\eeq

\section{\texorpdfstring{$G_2$}{G2} moduli space metric via dimensional reduction}
In this section we reduce $d=10$ heterotic string theory, correct to first order in $\ap$ on a $G_2$ manifold $Y$. In the first subsection we follow the results of \citeM applied to $G_2$ compactifications. The  identification of the Hull connection on $TY$ is central. In the second subsection we explore the reduction to $SU(3)$ as a consistency check of our results.

\subsection{ Reduction on \texorpdfstring{$Y$}{Y}, a \texorpdfstring{$G_2$}{G2} manifold}
 We start with a parameter space $\cM_{G_2}$ with coordinates $y^a$. For each point in $\cM_{G_2}$ we get a $G_2$ compactification satisfying the $N=1$ $d=3$ supersymmetry conditions and Bianchi identity. Equivalently, a section $\ccY_a$:
$$
\ccY_a \= 
\begin{pmatrix}
 M_a\\
 \aa_a
\end{pmatrix}~, \qquad \Dch \ccY_a \= 0~.
$$
That $\Dch^2 = 0$ means we have an underlying cohomology and there is a gauge fixing condition \eqref{eq:gaugefix}. Physically, this is enough to uniquely fix $\ccY_a$. We use this as our basis for a Kaluza-Klein expansion of supergravity.

We denote the $\IR^{2,1}$ coordinates by $X^e$, internal coordinates by $x^m$. The background field expansion of the metric, B-field and dilaton is
\beq\begin{split}
\dd s^2 &\= \big(g_{ef} + \d g_{ef}(X)\big)\, \dd X^e \otimes\dd X^f +
\big(g_{mn}(x) + \d g_{mn}(x,X)\big)\, \dd x^m\otimes \dd x^n~,\\[3pt]
B &\=  \d B_{ef}(X)\, \dd X^e  \dd X^f + (B_{mn}(x) + \d B_{mn}(x,X))\, \dd x^m  \dd x^n~,\\[3pt]
\Phi &\=  \phi_0+\vph(X) + \phi(x,X)~.
\end{split}\label{eq:ReductionAnsatz1}\eeq
The dilaton does not interest us. As explained in \citeAQS to cubic order in $\ap$, the non-trivial component on the internal manifold can be gauge fixed to be a constant. In other words $\phi(x,X)=\cO(\ap^3)$. The spacetime dilaton $\vph(X)$ decouples from the internal theory at this order in the $\ap$-expansion.  All small variations of background fields are regarded as quantum fluctuations.  

The decomposition of the 10D gauge field $A_M$ is partially fixed by the representation theory:
\beq
\text{Adj}(\Le_8) ~=~
(\text{Adj}(\Lg) ,{\rep 1}) \oplus_i (\rep{R}_i, \rep{r}_i)\oplus_i (\brep{R}_i,\brep{r}_i) \oplus ({\rep 1},\text{Adj}(\eu{G}))~,
\label{StandardAdjointDecomposition}\eeq
where if the vector bundle has structure group $\eu{G}$ then the spacetime gauge group is the commutant in $E_8$. We denote $\Lg$ the algebra for the unbroken gauge group $G = [\eu{G}, \Le_8]$.  Then,   $\rep{R}_i$ is the matter field representation and its conjugate $\brep{R}_i$ of $\Lg$. It is obviously possible the matter fields appear in real  or psuedo-real representations $\rep{R}_i=\brep{R}_i$. We take these terms to be captured by an element of the summand $(\rep{R}_i, \rep{r}_i)\oplus(\brep{R}_i,\brep{r}_i)$  with a slight abuse of notation. In any event, the matter fields are not relevant for our calculation here.

We now compute the metric explicitly by dimensionally reducing the supergravity action and identifying the coefficient of the kinetic terms of the moduli fields. We start with the Lagrangians arising from the Ricci-scalar $\cL_g$ and H-field strength $\cL_H$ before including the Yang-Mills term $\cL_F$ in subsequent subsections. An important role is played by the partner to the Yang-Mills term given by variations of the Hull connection $\cL_K$. 
  
The expansion of the Ricci scalar to quadratic order is
\beq
\begin{split}
  \pre{10}\cR ~=  \pre{3}\cR &- \pre{3}\nabla^2 \log \det (g_7+\d g_7) -
\qrt \Big(\pre{3}\nabla \log \det (g_7 + \d g_7) \Big)^2 \cr
&- \qrt \,g^{mn}g^{pq}\del_e( \d g_{mp})\,\del^e (\d g_{nq}) + \cdots~.
\label{ricciscalar}\end{split}
\eeq
Here $\pre{3}\cR$ is the $d=3$ Ricci scalar for the metric 
$\dd s_{(3)}^2 = (g_{ef}+\d g_{ef}(X))\, \dd X^e\otimes \dd X^f$,  
while the last term will give rise to the moduli space metric. The first three terms in \eqref{ricciscalar} recombine into the $d=3$ Ricci scalar after changing to Einstein frame. To see this, we need to include the $d=3$ dilaton field defined by
$$
e^{-2\phi_3(X)} ~=~ \frac{ e^{-2\varphi(X)}}{g_s^{2}V_0} \int_Y \! d^7 x \sqrt{g_7+\d g_7} ~,
$$ 
where $g_s = e^{\phi_0}$ is the zero-mode of the dilaton.
Then,  a Weyl transformation on the $d=3$ metric $g_{E\, ef} = e^{-2\phi_3} g_{ef}$ collapses the first three terms in  \eqref{ricciscalar} into the Einstein frame Ricci scalar. The ten-dimensional action \eqref{eq:10daction} under dimensional reduction gives 
\beq
\begin{split}
S ~=~  \frac{g_s^2}{2\k_3^2}\int d^4X \sqrt{g_{E}} \left( \pre{3}\cR_E  +\cL \right)~,
 \end{split}\label{MetricReduct}
\eeq
where  $\k_{10}^2 = V_Y g_s^{-2}\k_3^2$, $V_Y$ is the volume of the $G_2$ manifold and $\cL = \cL_g + \cL_B + \cL_S + \cL_F$ is the three-dimensional Lagrangian for the kinetic terms of the moduli fields coming from the reduction of the first four terms of \eqref{eq:10daction}.  We compute each of these terms below. The remaining terms in \eqref{eq:10daction} are at least $\cO(\ap^2)$ and are ignored. 

The first term  $\cL_g$ comes from the $\cR$ in \eqref{ricciscalar} and is given by
 \beq
\begin{split}
 \cL_g ~&= - \frac{1}{4V_Y} \int_Y d^7x \sqrt{g}\, g^{mn} g^{pq}\, \del_e y^a (\d_a g_{mp})\, \del^e y^b (\d_b g_{nq})~.
 \end{split}
\notag\eeq
Here $\d g_{mn} \cong \d y^a \d_a g_{mn}$ corresponds to a field variation and writing the space of operators as $\d_a$ acting on the background fields.  We also use the notation $\del_e y^a \cong \del_e (\d y^a) $ so that our expressions remain semi-compact. 

The next term $\cL_B$ comes from  the kinetic term for $B$:
\beq
\label{SH}
\cL_B ~= -  \frac{1}{2V_Y }\int_Y \dd^7 x \sqrt{g}\,  |H+\d H|^2~.
\eeq
In this expression, $H$ has all three legs along $Y$, while $\d H$ always has a leg in three-dimensional spacetime, so $H\star \d H = 0$ leaving 
\beq
\begin{split}
 \cL_B ~= -  \frac{1}{4V_Y }\int_Y \dd^7 x \sqrt{g} \, \d H_{emn}  \d H^{emn} ~.
\end{split}
\notag\eeq
Now, by Poincare invariance of $\IR^{2,1}$ the only relevant term to the metric in $\d H$ is one with a spacetime derivative of $\d y^a$:
\beq\notag
\d H_{emn}  \=  (\del_e y^a)  \ccB_{a\,mn} ~,
\eeq

Putting these together we find
\beq\label{eq:cLgb}
\begin{split}
 \cL_g + \cL_B &\= - \frac{1}{4V_Y} \int_Y \vol \left\{  \left(\d_a g_{mn} \d_b g^{mn} + \ccB_{a\,mn} \ccB_b{}^{mn} \right)  \del_e y^a \del^e y^b\right\} \\
&\= - \frac{1}{V_Y} \int_Y \vol \left\{  M_{a\,mn} M_b{}^{mn} \del_e y^a \del^e y^b\right\}  ~,
\end{split}
\eeq
where we use that $\ccB_a$ is antisymmetric, $\d_a g$ is symmetric and
\beq\label{eq:defM}
M_{a\,mn} \= \half \left(\d_a g_{mn} + \ccB_{a\,mn}\right)~.
\eeq

The Yang-Mills term is
\beq\notag
\cL_{F}\=\! -\frac{\ap}{4V_Y}\int_Y \vol \, \tr |\d F|^2 ~, \qquad |\d F|^2 \= \frac{1}{2} \d F_{MN} \d F^{MN}~.
\eeq
The gauge connection $A$ depends on parameters
$$
\d A \= \d y^a \aa_{a\,m} \dd x^m~, 
$$
and as above by promoting the parameters to dynamical $d=3$ moduli fields we find an effective lagrangian for the moduli
\beq
\begin{split}
\tr |F+\d F|^2 &\=  \tr\big( \d F_{e m} \d F^{e m}\big) +\cdots \= (\del_e y^a)( \del^e y^{b})\,\tr\big( \aa_{a\,m}\, \aa_b{}^m  \big) + \cdots ~.
\end{split}\notag
\eeq
There are omitted terms and these are ones that do not contribute the moduli space metric. The action is
$$
\cL_F \=\! -\frac{\ap}{4V_Y} \int_Y \vol\, \tr \left(\aa_{a\,m} \aa_b{}^m \right) \del_e y^a \del^e y^b~.
$$

We need to include the contribution of the deformations of the Hull connection on the tangent bundle. They appear formally in the same way as the Yang-Mills term with an opposite overall sign:
\beq\label{eq:cLK}
\cL_K \=  \frac{\ap}{4V_Y} \int_Y \vol\,\tr  \left(K_{a\,m} K_b{}^m \right) \del_e y^a \del^e y^b~.
\eeq
Our job is to solve this term by eliminating $K$ using \eqref{eq:dThH}. 
Using that $\tr K_{a\,q} K_b{}^q = g^{qp}\,K_{a\,q}{}^s{}_t K_{b\,p}{}^t{}_s $, together with \eqref{eq:dThH}, we get
\beq\notag
\begin{split}
 \tr K_{a\,q} K_b{}^q &\= 
- (\nabla_s M_{a\,qt} - \nabla_t M_{a\,qs} )(\nabla^s M_b{}^{qt} - \nabla^t M_b{}^{qs} ) ~.
\end{split}
\eeq
Hence, putting this into $\cL_K$ and integrating by parts
\beq\label{eq:LK1}
\begin{split}
\cL_K &\=\frac{\ap}{4V_Y} \int_Y \vol\,\tr \left(K_{a\,q} K_{b}{}^q\right) \del_e y^a \del^e y^b \\
&\= \frac{\ap}{2V_Y} \int_Y \vol \Big\{  M_{a qt} (\nabla_s\nabla^s M_b{}^{qt} - \nabla_s\nabla^t M_b{}^{qs})   \Big\} \del_e y^a \del^e y^b~.
\end{split}
\eeq
We now use a few little results. First, the commutator
\beq\label{eq:commutatorM}
[\nabla_s, \nabla_t] M_b{}^{qs} \= R_{st}{}^q{}_p M_b{}^{ps} + R_{st}{}^s{}_p M_b{}^{qs} \= R_{st}{}^q{}_p M_b{}^{ps}~,
\eeq
where because $\nabla$ is the Levi-Civita connection to this order in $\ap$ the Ricci-tensor $R_{st}{}^s{}_p = \cO(\ap)$. There is a Lichnerowicz equation for the field $M_a$. To derive this, recall the action \eqref{eq:10daction}. The equation of motion, to zeroth order in $\ap$, for $\d B$ is given by
\beq\notag
\begin{split}
- \half(\dd^\dag \dd \d B)_{qr}   &\=  \frac{1}{3}  \left( \nabla^p \nabla_{p} \d B_{qr} + R^p{}_{qr}{}^s  \d B_{sp}  - R^p{}_{rq}{}^s \d B_{sp}  +  \nabla_{r} \nabla^p  \d B_{pq} -  \nabla_q \nabla^{p} \d B_{pr} \right) \= 0~.
\end{split}
\eeq
The equation of motion for the metric to zeroth order in $\ap$ comes from $R(g+\d g) = R(g) = 0$, which in the gauge $\nabla^p \d g_{pq} = \cO(\ap)$ is the Lichnerowicz equation. If we also impose the gauge $\nabla^p \d B_{pq} = \cO(\ap)$, which is consistent with \eqref{eq:gaugefix}, viz. $\nabla^p M_{pq} = \cO(\ap)$, then we find equations of motion:
\beq\notag
\begin{split}
\nabla^p \nabla_{p} \d B_{qr} - R^p{}_{qr}{}^s  \d B_{ps}  +R^p{}_{rq}{}^s \d B_{ps}   &\= 0 + \cO(\ap)~, \\[3pt]
  \nabla^p \nabla_p \d g_{qr} - 2 R^s{}_{qr}{}^p   \d g_{ps} &\=0 +  \cO(\ap)~.
\end{split}
\eeq
Adding these two together
\beq\label{eq:G2Lichnerowicz}
\begin{split}
&\nabla^p \nabla_{p} M_{qr} - 2R^p{}_{qr}{}^s  M_{ps}   \= 0~.
\end{split}
\eeq
Putting together \eqref{eq:commutatorM}, \eqref{eq:G2Lichnerowicz} with the gauge fixing $\nabla^p M_{pq} = \cO(\ap)$ we find
\beq\notag
\begin{split}
\cL_K  &\=- \frac{\ap}{2V_Y} \int_Y \vol \Big\{  M_{a\, qt} R^{qpts} M_{b\,ps}   \Big\} \del_e y^a \del^e y^b~.
\end{split}
\eeq

To finish the calculation, we put together the above results 
$$
\cL \= \cL_g + \cL_B  + \cL_F + \cL_K~,
$$
to give
\beq\notag
\cL \=\! - \frac{1}{V_Y} \int_Y \vol \left(   M_{a\,mn} M_b{}^{mn} + \frac{\ap}{2} M_{a\, mn} R^{mpnq} M_{b\,pq}  +\frac{\ap}{4}  \tr \left(\aa_{a\,m} \aa_b{}^m \right)  \right) \del_e y^a \del^e y^b + \cO(\ap^2)~.
\eeq
We identify the $G_2$ moduli space metric from the kinetic terms of the moduli fields
\beq
\label{eq:G2modulimetric}
 \!\!\!\!G_{ab} \=  \frac{1}{V_Y} \int_Y \vol \left(   M_{a\,mn} M_b{}^{mn} + \frac{\ap}{2} M_{a\, mn} R^{mpnq} M_{b\,pq}  +\frac{\ap}{4}  \tr \left(\aa_{a\,m} \aa_b{}^m \right)  \right)  + \cO(\ap^2)~.
\eeq
This is  the $G_2$ moduli space metric conjectured in \citeMSS. This proves it as at the moduli space metric for the physical heterotic theory. The role of $\ap$-corrections and eliminating the spurious modes were important.   

As an aside, we can factorise the first two terms via a field redefinition $\wt M_{a\,mn} = M_{a\,mn} + \frac{\ap}{4} M_{a\,mn} R_{mpnq} M_a{}^{pq}$ rendering  a Weyl--Peterson inner product. However, the field $\wt M_a$ does not transform in the same was $M_{a}$ under diffeomorphisms due to the $\ap$-correction -- it is not of the form of $\d g + \ccB$. So the metric and its $\ap$-correction is non-trivial. A second point is the observation that if  $\cL_K$ is evaluated in the coordinate frame, using \eqref{eq:KinM}, then $\cL_K =0$. 

Returning to \eqref{eq:G2modulimetric}, we can dimensionally reduce to $SU(3)$ by an ansatz $Y= X \times S^1$, where $X$ is an $SU(3)$ manifold and the circle decompactifies to enhance $\IR^{2,1}\to \IR^{3,1}$. The fields reduce as
\beq \label{eq:MtoX}
\begin{split}
\d y^a   M_{a\,\rb \mb} &\= \d y^a \D_{a\,\rb \mb} \= \d y^\a \D_{\a\,\rb \mb}~,\\
\d y^a  M_{a\,\sb \m} &\= \half\d y^a   \ccZ_{a\,\sb \m} \=  - \half \d y^\a \ccZ_{\a\, \m\sb} ~,
\end{split}
\eeq
where on the left hand side we have written coordinates on the $G_2$ moduli space and on the right hand side these are embedded into the $SU(3)$ moduli space coordinates. The field $M$ is real and so this equation is to be supplemented by its complex conjugates. We have introduced holomorphic indices and note that the $SU(3)$ holomorphic gauge fixes $\ccZ_{\ab\m\sb} = 0$ and $\D_{\ab\mb\nb} = 0$. This leaves us with the right hand side. 

We now substitute this into the $G_2$ moduli space metric and we find
\beq\notag
\begin{split}
 G\#_{\a\bb} &\= \frac{1}{V_X}\int_X \Big( \D_\a{}^\m\star\D_{\bb}{}^{\nb}\,g_{\m\nb} +\qrt  \ccZ_\a^{(1,1)}\star\ccZb_{\bb}^{(1,1)} + \frac{\ap}{4}\tr{ ( \fD_\a A \star \fD_{\bb}A ) } \Big)\\[0.1cm]
 &\hspace{3cm}~+~ \frac{\ap}{2V_X}\int_X \vol ~ \left( \D_{\a\mb\nb}\,\D_{\bb\r\s} +\qrt \ccZ_{\a\, \r\mb}\,\ccZb_{\bb\,\s\nb} \right) \, R^{\mb\r\nb\s} + \cO(\ap^2)~.
\end{split}\raisetag{1.7cm}\eeq
This reduction result was evidence for the conjectured moduli space metric in \citeMSS. We have now proved this at the level of heterotic supergravity, including the first $\ap$-correction.

\section{Conclusion}
In this paper, we have derived the moduli space metric of heterotic $G_2$ compactifications from the ten-dimensional heterotic supergravity at ${\cal O(\alpha')}$, confirming the result conjectured in \cite{mcorist2024physicalmoduliheteroticg2}. This metric allows the construction of an adjoint operator $\Dch^\dag$, whose kernel describes a natural gauge fixing for the moduli, as discussed in \citeMSS.

\subsection*{Acknowledgements}
JM is in part supported by an ARC Discovery Project Grant DP240101409. We would like to thank MATRIX institute for hospitality where some of this work was completed.

\appendix

 \section{Converting between Lorentz and coordinate indices on connections}
 \label{s:AppSpin}
 The Levi--Civita connection in coordinate indices is given by the Christoffel symbols. Its deformation is 
 \beq\notag
 \d \G_m{}^p{}_q \= g^{sp} (\nabla_{[q} \d g_{s]m} + \half \nabla_m \d g_{sq}). 
 \eeq
Lower the middle index using the metric
 \beq\label{eq:dG}
 \d \G_{m\,pq} \=(\nabla_{[q} \d g_{p]m} + \half \nabla_m \d g_{pq}). 
 \eeq
We want to convert this to a deformation of the spin connection, $\d \Th_{ab}$ and metric compatibility requires antisymmetry in $ab$. A priori, $\d \G$ is not antisymmetric in $pq$. 

First recall the vielbein
\beq\notag
g_{pq} \= \eta_{ab} e^a_p e^b_q ~, \qquad\qquad \eta_{ab} \= E^p_a E^q_b g_{pq} ~, \qquad e^a = e^a_p \dd x^p~.
\eeq
Here $\eta_{ab} = {\rm diag}(1,\cdots,1)$. 
Taking a small deformation we find
\beq\label{eq:defgvielbein}
\begin{split}
  \d g_{pq} \= \eta_{ab} (\d e^a_p e^b_q +  e^a_p \d e^b_q) \= \d e^a_p e_{aq} + e_{ap} \d e^a_q ~.
\end{split}
\eeq
A small abuse of notation $e_{aq} = \eta_{ab} e^b_q$ so that $\d e_{aq} = \eta_{ab} \d e^b_q$. It is also true that 
$$
\d e_{aq}  = \half \d g_{qp} E^p_a~,
$$
which can be checked by substituting  back into \eqref{eq:defgvielbein}. 

The relationship between the spin connection $\Th$ and the Christoeffel symbols $\G$ is in say \cite{Eguchi:1980jx}:
\beq\notag
\begin{split}
 \Th_{m\,ab} &\= E_a^p E_b^q \G_{m\,pq} -  E_b^n \del_m e_{an}~.
\end{split}
\eeq

A small deformation is
\beq\notag
\begin{split}
\d \Th_{m\,ab}&\= E_a^p E_b^q \d \G_{m\,pq} -  E_b^q \nabla_m \d e_{aq}  -  \d E_b^q     \nabla_m e_{aq} ~.
\end{split}
\eeq
Substitute into \eqref{eq:dG},
\beq\notag
\d \Th_{m\,ab} \= E_a^p E_b^q \left(\nabla_{[q} \d g_{p]m} + \half \nabla_m \d g_{pq} \right)        - ( \nabla_m\d e_{aq} ) E_b^q     -  (\nabla_m e_{aq}) \d E_b^q ~.
\eeq
Using that $\d e_{aq}  = \half \d g_{qp} E^p_a$, 
\beq\label{eq:dThdGam}
\begin{split}
 \d \Th_{m\,ab} &= E_a^p E_b^q \left(\nabla_{[q} \d g_{p]m} {+} \half \nabla_m \d g_{pq} \right)        {-} \half ( \nabla_m  \d g_{qp} ) E^p_aE_b^q  {-} \half ( \nabla_m  E^p_a )\d g_{qp} E_b^q     {-}  (\nabla_m e_{aq}) \d E_b^q \\
 &= E_a^p E_b^q \left(\nabla_{[q} \d g_{p]m}  \right)         -  ( \nabla_m  E^p_a ) \d e_{bp}     -  (\nabla_m e_{aq}) \d E_b^q \\
  &= E_a^p E_b^q \left(\nabla_{[q} \d g_{p]m}\right)~.
\end{split}\raisetag{1cm}
\eeq
To see the last equality we used metric compatibility of the vielbein: 
$$
\nabla_m e^a_p \=\! -\Th_m{}^a{}_c e^c_p~, \qquad \nabla_m E^p_a \= \Th_m{}^c{}_a E^p_c~, \quad \Th_{mab} \=\! -\Th_{mba}~.
$$
Hence, 
$$
\d e_{bp} \nabla_m E_a^p \= -\Th_{mac} e^c_q \d e_{bp} g^{pq}~, \qquad \d E^q_b \nabla_m e_{aq} = \Th_{mac} e^c_q g^{qp} \d e_{bp}~.
$$
To conclude, we write down the deformation of the Hull connection in the spin frame:
\beq\label{eq:ThH2}
 \d \Th^\H_{m\,ab}  \= E_a^p E_b^q \left(\nabla_{[q} \d g_{m|p]} + \nabla_{[q} \ccB_{m|p]}\right) \=  2E_a^p E_b^q\, \nabla_{[q} M_{m|p]}~.
\eeq
We read off how this transforms under small gauge
$$
 \d \Th^\H_{m\,ab} \sim  \d \Th^\H_{m\,ab} - R_{mabq} \ve^q + \nabla_m (\nabla_b \ve_a - \nabla_a \ve_b)~.
$$
From the fact this is a gauge connection with a canonical transformation law \\$ {\d \Th^\H{}_{ab} \sim  \d \Th^\H{}_{ab} - R_{abq} \ve^q+   \dd \psi_{ab}}$ we identify $\psi_{ab} = - (\nabla_a \ve_b - \nabla_b \ve_a)$.


\bibliographystyle{utphys}
\providecommand{\href}[2]{#2}\begingroup\raggedright\endgroup

\end{document}